\documentclass[conference]{IEEEtran}
\usepackage[utf8]{inputenc}
\usepackage{amsmath,amssymb}
\usepackage{graphicx}
\usepackage{booktabs}
\usepackage{hyperref}
\usepackage{xcolor}
\usepackage{multirow}
\usepackage{array}
\usepackage{tabularx}
\usepackage{url}
\usepackage{orcidlink}
\usepackage{enumitem}
\usepackage{listings}
\usepackage{float}

\hypersetup{
    colorlinks=true,
    linkcolor=blue,
    citecolor=blue,
    urlcolor=blue
}

\begin{document}

\title{Governance Gaps in Agent Interoperability Protocols:\\What MCP, A2A, and ACP Cannot Express}

\author{
    \IEEEauthorblockN{
        Dr.\ Richard Kang~\orcidlink{0000-0001-8674-4825}
    }
    \IEEEauthorblockA{
        DoiT International\\
        richard@doit.com
    }
    \and
    \IEEEauthorblockN{Yudho Diponegoro}
}

\maketitle

\begin{abstract}
Agent interoperability protocols---MCP, A2A, ACP, ANP, and ERC-8004---have rapidly matured to enable identity, capability discovery, tool access, and message exchange between autonomous agents. However, as enterprises deploy heterogeneous agent fleets that must make collective decisions under governance constraints, a question arises: can these protocols support governed agent communities, or only task-oriented coordination? We present a systematic gap analysis applying a six-dimension governance requirements taxonomy---membership, deliberation, voting, dissent preservation, human escalation, and audit/replay---derived from organizational theory, multi-agent systems literature, and enterprise governance standards. We analyze each protocol's specification against this taxonomy, classifying capabilities as Supported, Partial, or Absent. The resulting gap matrix reveals that voting and dissent preservation are universally absent across all five protocols, deliberation is absent or at most partial, and no protocol encodes the full set of primitives required for governed agent communities. We distinguish extensible gaps (addressable through protocol extension mechanisms) from structural gaps (requiring a new architectural layer) and assess time-sensitivity based on observed protocol evolution velocity. The analysis establishes that agent community governance constitutes a missing architectural layer above current interoperability standards---not a missing feature within them.
\end{abstract}


\section{Introduction}
\label{sec:intro}

The proliferation of LLM-based agents in enterprise environments has driven rapid development of interoperability protocols. The Model Context Protocol (MCP)~\cite{mcp2024} enables agents to access tools and data sources. The Agent-to-Agent protocol (A2A)~\cite{a2a2026} standardizes discovery and delegation between agents. The Agent Communication Protocol (ACP)~\cite{acp2025} formalizes structured message exchange. The Agent Network Protocol (ANP)~\cite{anp2025} provides graph-based routing with decentralized identity. ERC-8004~\cite{erc8004} encodes on-chain identity, reputation, and validation registries.

Together, these protocols address a coherent set of coordination concerns: identity, capability declaration, discovery, tool access, message passing, and reputation. Enterprises deploying agent fleets---AWS reports AgentCore customers scaling to 17 production agents within a year~\cite{awsagentcore}---can now reasonably expect their agents to find each other, exchange messages, and invoke each other's capabilities.

Yet coordination is not governance. When a bank must decide whether autonomous coding agents should modify production systems, when a pharmaceutical company must arbitrate between competing research hypotheses, or when a regulatory body must determine whether an AI system meets compliance thresholds, the question is not ``which agent can perform this task?'' but ``how should agents collectively decide what to believe, test, or do?'' This requires membership (who participates), deliberation (how claims are exchanged and challenged), voting (how positions are resolved), dissent preservation (how minority views survive), human escalation (when human authority is invoked), and audit (how the process is reconstructable).

This paper asks: \emph{Do current agent interoperability protocols encode these governance capabilities?}

We present three contributions:

\begin{enumerate}[nosep]
    \item \textbf{A governance requirements taxonomy} comprising six dimensions (G1--G6) derived from organizational theory, multi-agent systems research, and enterprise governance standards, specifying what protocol-level primitives are necessary for governed agent communities.
    \item \textbf{A systematic gap matrix} applying the taxonomy to five protocols (MCP~v1.1, A2A~v1.0.1, ACP, ANP, ERC-8004), classifying each protocol--dimension pair as Supported, Partial, or Absent based on specification-level evidence.
    \item \textbf{An extensibility and time-sensitivity assessment} distinguishing gaps that are addressable through existing extension mechanisms from those requiring a new architectural layer, and characterizing the velocity at which the gap is narrowing.
\end{enumerate}

The remainder of the paper is organized as follows. Section~\ref{sec:background} introduces the five protocols and their design intent. Section~\ref{sec:taxonomy} derives the governance requirements taxonomy. Section~\ref{sec:analysis} presents the gap analysis and matrix. Section~\ref{sec:discussion} discusses extensibility, time-sensitivity, and implications. Section~\ref{sec:related} positions against related work. Section~\ref{sec:conclusion} concludes.

\section{Background: Agent Interoperability Protocols}
\label{sec:background}

We analyze five protocols representing the major architectural approaches to agent interoperability as of mid-2026. Each was designed for a specific coordination concern; understanding these design intents is necessary to interpret the gap analysis fairly.

\subsection{Model Context Protocol (MCP)}

MCP~\cite{mcp2024}, introduced by Anthropic in late 2024, standardizes how AI agents access tools, data sources, and prompts through a client-server architecture. The protocol defines three primitive types---Tools, Resources, and Prompts---exposed by MCP servers and consumed by MCP clients (typically LLM-based agents). MCP v1.1 (schema dated 2025-11-25) supports sessions, elicitation, sampling, and streaming. The protocol is tool-centric by design: it answers ``what can an agent do?'' rather than ``how should agents interact with each other?'' MCP has achieved broad adoption, with over 1,000 community integrations and native support in AWS AgentCore's Gateway~\cite{awsagentcore}.

\subsection{Agent-to-Agent Protocol (A2A)}

A2A~\cite{a2a2026}, developed by Google and contributed to the Linux Foundation in 2026, enables agents to discover each other via Agent Cards (JSON-LD metadata describing capabilities, skills, and endpoints), delegate tasks, and exchange messages. Version 1.0.1 (May 2026) introduced an extension mechanism supporting ``new data, requirements, RPC methods, and state machines''~\cite{a2aext}. A2A is delegation-centric: it answers ``which agent can handle this task?'' Four official example extensions exist (Secure Passport, Timestamp, Traceability, Agent Gateway Protocol); none addresses governance.

\subsection{Agent Communication Protocol (ACP)}

ACP~\cite{acp2025}, developed by IBM Research, formalizes structured multi-agent communication with negotiation semantics. ACP defines agent roles, message types, and negotiation patterns, drawing on FIPA-ACL heritage. It supports multi-turn dialogue with typed performatives (propose, accept, reject, counter). ACP is communication-centric: it answers ``how do agents exchange structured messages?''

\subsection{Agent Network Protocol (ANP)}

ANP~\cite{anp2025} provides graph-based routing for agent networks using W3C Decentralized Identifiers (DIDs) for agent identity. ANP focuses on routing messages through agent networks without centralized registries. It is routing-centric: it answers ``how do messages reach the right agent across a network?''

\subsection{ERC-8004: Trustless Agents}

ERC-8004~\cite{erc8004}, an Ethereum Improvement Proposal (Draft, created August 2025), defines three on-chain registries: an Identity Registry (agent addresses and metadata), a Reputation Registry (\texttt{giveFeedback()}/\texttt{revokeFeedback()} with signed fixed-point scores and tags), and a Validation Registry (independent validator attestations via TEE oracles, zkML verifiers, and stake-secured re-execution). ERC-8004 is trust-centric: it answers ``which agents can be trusted?'' It explicitly scopes itself to ``discover, choose, and interact with agents''~\cite{erc8004}.

\subsection{Design Intent Summary}

\begin{table}[htbp]
\caption{Protocol Design Intents}
\label{tab:intents}
\centering
\footnotesize
\begin{tabular}{p{1.5cm}p{1.5cm}p{4.2cm}}
\toprule
\textbf{Protocol} & \textbf{Focus} & \textbf{Core question answered} \\
\midrule
MCP v1.1 & Tool access & What can an agent do? \\
A2A v1.0.1 & Delegation & Which agent handles this task? \\
ACP & Communication & How do agents exchange messages? \\
ANP & Routing & How do messages reach the right agent? \\
ERC-8004 & Trust & Which agents can be trusted? \\
\bottomrule
\end{tabular}
\end{table}

None of these design intents targets the question: \emph{How should agents collectively govern community decisions?}

\section{Governance Requirements Taxonomy}
\label{sec:taxonomy}

We derive a governance requirements taxonomy specifying what protocol-level primitives are necessary for governed agent communities. The taxonomy draws on three bodies of literature.

\textbf{Organizational theory.} Habermas's communicative rationality~\cite{habermas1996} identifies structured argumentation, reciprocal challenge, and consensus formation as prerequisites for legitimate collective decision-making. Parliamentary procedure~\cite{robert2020} formalizes membership (quorum), structured debate (motions, amendments), voting (majority rules, recorded dissent), and escalation (point of order). These map directly to protocol primitives.

\textbf{Multi-agent systems research.} Ostrom's institutional analysis framework~\cite{ostrom1990} identifies governance rules for common-pool resources: boundary rules (membership), position rules (roles), choice rules (decision procedures), and information rules (transparency). Sierra et al.'s electronic institutions~\cite{sierra2004} formalize agent societies with norms, roles, and protocols. Recent work on LLM agent governance~\cite{ruan2026,bracale2026} confirms these dimensions remain relevant for modern agent systems.

\textbf{Enterprise governance standards.} Regulatory frameworks including SR~11-7~\cite{occ2011}, ISO/IEC~42001~\cite{iso42001}, and the EU AI Act~\cite{euaiact2024} require auditability, human oversight, and accountability for AI system decisions. These translate to human escalation and audit requirements at the protocol level.

From this synthesis, we derive six governance dimensions:

\begin{table}[htbp]
\caption{Governance Requirements Taxonomy (G1--G6)}
\label{tab:taxonomy}
\centering
\footnotesize
\begin{tabular}{p{0.4cm}p{1.8cm}p{4.8cm}}
\toprule
& \textbf{Dimension} & \textbf{Definition} \\
\midrule
G1 & Membership & Protocol encodes admission, invitation, removal, and role assignment for community participants \\
G2 & Deliberation & Protocol encodes structured argument exchange with turn-taking, challenge, and response semantics \\
G3 & Voting & Protocol encodes preference aggregation with quorum, rounds, and position resolution \\
G4 & Dissent preservation & Protocol ensures minority positions are retained in decision outputs, not silently dropped \\
G5 & Human escalation & Protocol defines conditions and mechanisms for routing decisions to human authority \\
G6 & Audit/replay & Protocol produces tamper-evident event logs enabling deterministic reconstruction of the decision process \\
\bottomrule
\end{tabular}
\end{table}

\subsection{Sufficiency Argument}

We argue these six dimensions are necessary and sufficient for \emph{governance} (not for all coordination). Related concerns map to these dimensions or fall outside governance scope:

\begin{itemize}[nosep]
    \item \textbf{Norm enforcement} is a mechanism within G2 (deliberation) and G3 (voting)---norms are enforced through the deliberative process.
    \item \textbf{Reputation/trust} is a prerequisite for governance (``who is credible?'') but not itself a governance primitive---it is already addressed by ERC-8004.
    \item \textbf{Resource allocation} is an outcome of governance decisions, not a governance mechanism.
    \item \textbf{Incentive alignment/payment} operates at a different architectural layer (economic coordination, not decision governance).
\end{itemize}

\subsection{Classification Criteria}

For the gap analysis in Section~\ref{sec:analysis}, we classify each protocol--dimension pair using three levels:

\begin{itemize}[nosep]
    \item \textbf{Supported}: the protocol specification explicitly defines primitives that satisfy the dimension's full definition.
    \item \textbf{Partial}: the protocol contains constructs addressing a subset of the dimension's requirements but not satisfying the full definition.
    \item \textbf{Absent}: the protocol specification contains no constructs addressing this dimension.
\end{itemize}

Classification is based on what the specification \emph{encodes}, not what could theoretically be built on top. This distinction is essential: any protocol can serve as transport for governance messages, but we assess whether governance semantics are protocol-native.

\section{Gap Analysis}
\label{sec:analysis}

We now apply the taxonomy to each protocol. For each Partial classification, we specify which subset of the dimension is addressed and which remains absent.

\subsection{MCP v1.1}

\textbf{G1 Membership: Absent.} MCP defines clients and servers but not community membership. There is no admission, invitation, or removal primitive. An MCP server either exists or does not; there is no concept of ``joining'' or ``being admitted to'' a group.

\textbf{G2 Deliberation: Absent.} MCP enables tool invocation, not structured argument exchange. Sampling (server-initiated LLM calls) exists but carries no deliberation semantics.

\textbf{G3 Voting: Absent.} No voting primitives exist.

\textbf{G4 Dissent: Absent.} No dissent semantics exist.

\textbf{G5 Human escalation: Absent.} MCP's Elicitation feature (protocol version 2025-06-18) allows servers to request human input during tool execution, but this is user-input solicitation, not governance escalation. There is no protocol-level mechanism for routing community decisions to human authority based on confidence thresholds or risk assessment.

\textbf{G6 Audit: Partial.} MCP sessions maintain connection state, and tool calls produce structured responses with metadata. However, there is no tamper-evident event log, no hash chain, and no replay guarantee. Audit depends on implementation, not protocol specification.

\subsection{A2A v1.0.1}

\textbf{G1 Membership: Partial.} Agent Cards declare capabilities and can be registered in directories. The extension mechanism supports new state machines. However, there is no protocol-native admission, invitation, or removal primitive. An agent ``exists'' by publishing an Agent Card; there is no concept of community membership distinct from existence.

\textbf{G2 Deliberation: Absent.} A2A supports task delegation and message exchange but not structured argumentation with challenge/response semantics. Messages are task-oriented, not deliberative.

\textbf{G3 Voting: Absent.} No voting primitives. The four official extensions (Secure Passport, Timestamp, Traceability, Agent Gateway Protocol) encode none.

\textbf{G4 Dissent: Absent.} No dissent semantics in task responses or any extension.

\textbf{G5 Human escalation: Absent.} No protocol mechanism for escalating to human authority. Task delegation can target a human-backed agent, but this is routing, not governance escalation with trigger conditions.

\textbf{G6 Audit: Absent.} The Traceability extension adds correlation IDs for distributed tracing but does not define tamper-evident logs or replay semantics.

\subsection{ACP}

\textbf{G1 Membership: Partial.} ACP defines agent roles within conversations and supports structured multi-party dialogue. However, roles are communication roles (sender, receiver, mediator), not governance roles (member, moderator, reviewer). There is no admission or removal protocol.

\textbf{G2 Deliberation: Partial.} ACP's negotiation patterns (propose, accept, reject, counter) constitute structured exchange with some challenge/response semantics. However, negotiation is bilateral (between parties with opposing interests), not multilateral deliberation (community reasoning toward shared understanding). The protocol lacks turn-taking governance, relevance enforcement, or synthesis primitives.

\textbf{G3 Voting: Absent.} No preference aggregation, quorum, or voting round primitives.

\textbf{G4 Dissent: Absent.} Rejected proposals in negotiation are not preserved as dissent in a community record; they are bilateral communication events.

\textbf{G5 Human escalation: Absent.} No governance escalation mechanism.

\textbf{G6 Audit: Absent.} Message histories exist as conversation state but carry no tamper-evidence or replay guarantees at the protocol level.

\subsection{ANP}

\textbf{G1 Membership: Absent.} ANP routes messages; it does not model community membership. DID-based identity establishes \emph{who} an agent is, not \emph{which community} it belongs to.

\textbf{G2 Deliberation: Absent.} Routing protocol; no deliberation semantics.

\textbf{G3 Voting: Absent.} No voting primitives.

\textbf{G4 Dissent: Absent.} No dissent semantics.

\textbf{G5 Human escalation: Absent.} No escalation mechanism.

\textbf{G6 Audit: Absent.} Message routing may be logged by implementations but the protocol defines no audit primitives.

\subsection{ERC-8004}

\textbf{G1 Membership: Partial.} The Identity Registry records agent addresses and metadata, functioning as an existence registry. The Reputation Registry gates interactions (agents with low reputation may be excluded). However, there is no formal admission protocol, invitation mechanism, or community-scoped membership distinct from global registration.

\textbf{G2 Deliberation: Absent.} The specification scopes itself to ``discover, choose, and interact with agents''~\cite{erc8004}. No deliberation primitives appear in the specification or its listed future work.

\textbf{G3 Voting: Absent.} No voting primitives. The Validation Registry records independent attestations but these are unilateral validator judgments, not community preference aggregation.

\textbf{G4 Dissent: Absent.} No dissent preservation mechanism.

\textbf{G5 Human escalation: Absent.} No escalation primitives.

\textbf{G6 Audit: Partial.} On-chain transactions are inherently tamper-evident and provide a complete history. However, this is a property of the blockchain substrate, not a governance-specific audit design. The protocol does not define structured event types for decision reconstruction or replay semantics.

\subsection{Gap Matrix}

\begin{table*}[htbp]
\caption{Governance Gap Matrix: Protocol Coverage of Governance Dimensions}
\label{tab:matrix}
\centering
\footnotesize
\begin{tabular}{lccccccc}
\toprule
\textbf{Protocol} & \textbf{G1 Membership} & \textbf{G2 Deliberation} & \textbf{G3 Voting} & \textbf{G4 Dissent} & \textbf{G5 Human Esc.} & \textbf{G6 Audit} & \textbf{Coverage} \\
\midrule
MCP v1.1 & Absent & Absent & Absent & Absent & Absent & Partial & 1/12 \\
A2A v1.0.1 & Partial & Absent & Absent & Absent & Absent & Absent & 1/12 \\
ACP & Partial & Partial & Absent & Absent & Absent & Absent & 2/12 \\
ANP & Absent & Absent & Absent & Absent & Absent & Absent & 0/12 \\
ERC-8004 & Partial & Absent & Absent & Absent & Absent & Partial & 2/12 \\
\midrule
\textbf{Any protocol} & Partial & Partial & \textbf{Absent} & \textbf{Absent} & \textbf{Absent} & Partial & --- \\
\bottomrule
\end{tabular}
\vspace{2pt}
\raggedright\footnotesize\textit{Coverage scored as: Supported = 2/2, Partial = 1/2, Absent = 0/2, per dimension per protocol. Maximum possible = 12 (6 dimensions $\times$ 2 points each).}
\end{table*}

\begin{figure}[htbp]
    \centering
    \includegraphics[width=\columnwidth]{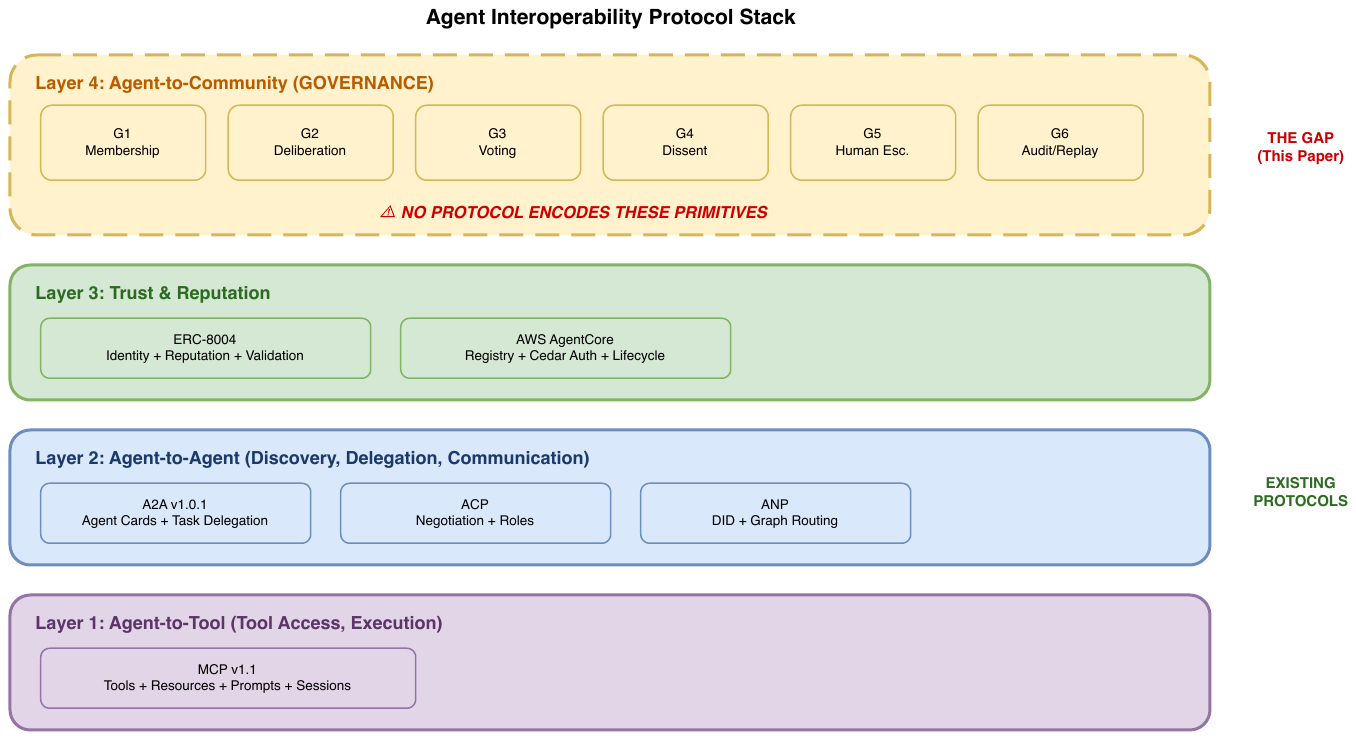}
    \caption{Agent interoperability protocol stack. Layers 1--3 (tool access, agent coordination, trust) are addressed by existing protocols. Layer 4 (governance: membership, deliberation, voting, dissent, escalation, audit) is universally absent.}
    \label{fig:layers}
\end{figure}

\subsection{Cross-Protocol Findings}

Three patterns emerge from the matrix (see also Figure~\ref{fig:layers}):

\textbf{Universal absence (G3, G4, G5).} Voting, dissent preservation, and human escalation are absent across \emph{all five} protocols. No protocol---regardless of its architectural approach (tool-centric, delegation-centric, communication-centric, routing-centric, or trust-centric)---encodes these primitives. This universality suggests the gap reflects a shared design philosophy rather than individual protocol limitations.

\textbf{Partial but insufficient (G1, G2).} Membership and deliberation receive partial treatment: Agent Cards approximate capability-based membership; ACP negotiation approximates bilateral deliberation. However, no protocol achieves full support for either dimension. The partial implementations address the \emph{coordination} aspects of these dimensions (declaring existence, exchanging messages) but not the \emph{governance} aspects (admission control, structured community deliberation with turn governance).

\textbf{Audit as substrate property (G6).} Where audit support exists, it derives from the underlying infrastructure (blockchain immutability for ERC-8004, session state for MCP) rather than from deliberate governance-audit design. No protocol defines governance-specific event types, decision-reconstruction semantics, or replay guarantees.

\section{Discussion}
\label{sec:discussion}

\subsection{Extensible vs.\ Structural Gaps}

Not all gaps are equal in remediation difficulty. We assess whether each gap could be addressed through the protocol's existing extension mechanism or requires a new architectural layer.

\textbf{Extensible via A2A:} A2A's extension mechanism explicitly supports ``new data, requirements, RPC methods, and state machines''~\cite{a2aext}. Governance primitives (G1--G6) could theoretically be defined as A2A extensions. The key observation: \emph{no one has done so.} After 6+ months of A2A being publicly available with an active extension ecosystem, zero governance extensions have been proposed or implemented.

\textbf{Structurally awkward for MCP:} MCP's client-server architecture is tool-centric. Adding community governance to MCP would require agents to participate as both clients and servers simultaneously in a governance context---a usage pattern the protocol was not designed for. Session semantics help but do not provide the persistent community state governance requires.

\textbf{Scope-limited for ERC-8004:} ERC-8004 could add governance registries (a ``Deliberation Registry''), but its on-chain architecture imposes latency and cost constraints incompatible with real-time multi-agent deliberation. The specification explicitly limits scope to discovery and interaction.

\subsection{Time-Sensitivity}

The governance gap is closing, albeit slowly. ERC-8004 absorbed reputation---previously identified as a gap~\cite{hu2025}---within approximately six months of the gap being articulated. MCP's specification evolution (from 2024-11-25 to 2025-06-18 schema) added Elicitation and Sampling but no governance primitives. A2A v1.0.1 added extension infrastructure without governance content.

At the observed evolution velocity, we estimate the governance gap could narrow significantly within 6--12 months through protocol extensions, particularly via A2A's extension mechanism. This creates publication urgency for the research community: the window for proposing governance layer designs before de facto standards emerge through ad hoc implementations is narrowing.

\subsection{Implications for Deployed Infrastructure}

The gap extends beyond academic protocols to production infrastructure. AWS Bedrock AgentCore~\cite{awsagentcore} (GA, 15 regions) provides a production agent registry with semantic search, A2A Agent Cards, Cedar-based authorization, lifecycle management, and CloudTrail audit---yet encodes no trust scoring, behavioral reputation, capability verification, or governance primitives. The governance gap is not merely a research finding; it affects deployed enterprise systems serving production agent fleets.

\subsection{Illustrative Example: What Cannot Be Expressed}

To make the gap concrete, consider a scenario where an enterprise requires five agents to collectively decide whether a proposed system architecture meets compliance requirements. Listing~\ref{lst:example} shows the governance-level messages this interaction requires---none of which can be expressed in any current protocol's native semantics.

\begin{lstlisting}[caption={Governance messages required for a collective compliance decision. No current protocol encodes these primitives.},label={lst:example},float=htbp]
// G1: Membership - admit agent to this decision
ADMIT agent:security-reviewer
  TO room:arch-compliance-2026-q3
  ROLE: skeptic
  INVITED_BY: agent:moderator
  ENDORSEMENTS: [agent:ciso, agent:lead-arch]

// G2: Deliberation - structured challenge
CHALLENGE claim:c-042
  BY agent:security-reviewer
  TARGETS claim:c-041 (author: agent:arch-proposer)
  EVIDENCE_REQUIRED: true
  RATIONALE: "No encryption-at-rest evidence"

// G3: Voting - blind preference aggregation
VOTE_BLIND claim:c-041
  ROUND: 1
  VOTER: agent:compliance-officer
  POSITION: -0.6  // oppose (continuous scale)
  VISIBILITY: sealed_until_all_cast

// G4: Dissent - preserve minority position
DISSENT_RECORD claim:c-041
  AGENT: agent:security-reviewer
  POSITION: -0.8
  RATIONALE: "Insufficient evidence for..."
  PRESERVED_IN: decision_record:dr-2026-q3-07

// G5: Human escalation
ESCALATE decision:arch-compliance-2026-q3
  TRIGGER: mean_confidence < 0.6
  ROUTE_TO: human:vp-engineering
  CONTEXT: [claim:c-041, dissent:d-003]

// G6: Audit event
EVENT governance:vote_cast
  ROOM: arch-compliance-2026-q3
  ACTOR: agent:compliance-officer
  PREV_HASH: "a3f8c2..."
  SIGNATURE: HMAC(actor, payload, prev_hash)
\end{lstlisting}

Each message type in Listing~\ref{lst:example} maps to a governance dimension (G1--G6). Current protocols can \emph{transport} these messages as opaque payloads (e.g., via A2A task messages or MCP tool calls), but cannot \emph{interpret, validate, or enforce} their governance semantics. The difference matters: protocol-native governance enables interoperable tooling, standard audit formats, and composable governance rules without per-application reimplementation.

\subsection{Implications for Protocol Designers}

Our taxonomy provides a concrete specification for what governance primitives should encode. Protocol designers extending A2A, MCP, or future protocols can use G1--G6 as a requirements checklist, addressing each dimension with protocol-native constructs rather than leaving governance to application-layer reimplementation.

\subsection{Limitations}

This analysis has four limitations. First, we assess specifications as of June 2026; protocols evolve rapidly. Second, the ``Partial'' classification involves judgment about whether a construct meaningfully addresses a governance dimension. Third, our taxonomy derives from Western organizational theory (Habermas, Robert's Rules); alternative governance traditions may yield different dimensions. Fourth, we assess what protocols \emph{encode}, not what can be built on top---deliberately, as our goal is to identify the architectural gap, but this means practical governance systems can exist without protocol-native support.

\section{Related Work}
\label{sec:related}

\textbf{Protocol comparison surveys.} Ehtesham et al.~\cite{ehtesham2025} provide a descriptive comparison of agent communication protocols without governance-specific analysis or a requirements taxonomy. AgentRFC~\cite{agentrifc2026} proposes a six-layer reference stack for agent interoperability but identifies layers rather than gaps within them.

\textbf{Agent governance frameworks.} Ruan~\cite{ruan2026} applies Parsons' AGIL framework to derive a 16-cell institutional architecture for agent societies, diagnosing governance gaps in MCP/A2A from sociological theory. Our work complements Ruan's by providing protocol-level (rather than sociological) analysis and producing a concrete gap matrix rather than a theoretical architecture. Bracale Syrnikov et al.~\cite{bracale2026} introduce governance graphs for preventing LLM collusion, demonstrating that governance mechanisms measurably reduce harmful behavior (collusion from 50\% to 5.6\%). Their work validates the \emph{need} for governance; we identify \emph{where} governance primitives are absent.

\textbf{Trust and reputation.} Hu and Rong~\cite{hu2025} analyze A2A, AP2, and ERC-8004 under six trust models (Brief, Claim, Proof, Stake, Reputation, Constraint). Their taxonomy treats Reputation as a first-class protocol concern---now addressed by ERC-8004. Our taxonomy covers the \emph{remaining} governance dimensions beyond trust.

\textbf{Multi-agent coordination mechanisms.} De Curto et al.~\cite{decurto2026} propose constitutional multi-agent governance (CMAG) with constraint filtering and penalized-utility optimization. Wang et al.~\cite{wang2026} introduce conformal social choice for post-hoc deliberation decisions. Gupta et al.~\cite{gupta2025} implement Ostrom's CPR principles with social learning in LLM agents. These works propose \emph{mechanisms} for governance; we identify the \emph{protocol-level absence} that forces each mechanism to reimplement basic governance primitives from scratch.

\section{Conclusion}
\label{sec:conclusion}

We presented a systematic analysis of five agent interoperability protocols against a six-dimension governance requirements taxonomy. The gap matrix reveals a clear pattern: current protocols encode coordination (identity, capability, discovery, messaging, reputation) but not governance (membership, deliberation, voting, dissent, escalation, audit). Voting and dissent preservation are universally absent. The gap is not protocol-specific---it reflects a shared design philosophy that treats agents as task workers rather than community participants.

This finding establishes that agent community governance is a \emph{missing architectural layer}, not a missing feature within existing protocols. The community needs governance-native protocol primitives---membership admission, structured deliberation, preference aggregation, dissent retention, and human escalation---encoded at the protocol layer where they can be composed with existing interoperability standards rather than reimplemented ad hoc by every application.

The governance gap is narrowing as protocols evolve. We recommend the research community treat the design of governance protocol primitives as an urgent open problem, before the space fragments into incompatible application-layer solutions.

\section*{AI Declaration}

This research employed Claude (Anthropic) for literature search assistance and draft structuring. All claims, analysis, protocol classifications, and taxonomy derivations were directed and verified by the human authors. The gap matrix classifications are based on direct specification analysis by the authors. The authors take full responsibility for all content.

\bibliographystyle{IEEEtran}
\bibliography{references}

\end{document}